\documentclass[apj,onecolumn]{emulateapj}

\shorttitle{Luminosity Correlations for Gamma-Ray Bursts}
\shortauthors{Sultana et al.}

\begin{document}

\title{Luminosity correlations for gamma-ray bursts and implications \\
for their prompt and afterglow emission mechanisms}

\author{J. Sultana\altaffilmark{1}, D. Kazanas\altaffilmark{2} and K. Fukumura\altaffilmark{2}}

\email{joseph.sultana@um.edu.mt}

\altaffiltext{1}{Mathematics Department, Faculty of Science,
University of Malta, Msida MSD2080 Malta.}
\altaffiltext{2}{Astrophysics Science Division, NASA/Goddard Space
Flight Center, Greenbelt, MD 20771 USA.}

\begin{abstract}
We present the relation between the ($z-$ and $k-$corrected)
spectral lags, $\tau$, for the standard \textit{Swift} energy bands
$50-100$ keV and $100-200$ keV and the peak isotropic luminosity,
$L_{\mathrm{iso}}$ (a relation reported first by Norris et al.),
for a subset of $12$ long \textit{Swift} GRBs taken from a
recent study of this relation by Ukwatta et al. The chosen GRBs
are also a subset of the Dainotti et al. sample, a set of
{\em Swift} GRBs of known redshift, employed in establishing a
relation between the (GRB frame) luminosity, $L_X$, of the shallow
(or constant) flux portion of the typical XRT GRB-afterglow light
curve and the (GRB frame) time of transition to the normal decay
rate, $T_{\mathrm{brk}}$. We also {present} the
$L_X-T_{\mathrm{brk}}$ relation using only the bursts common in the
two samples. The two relations exhibit a significant degree of
correlation ($\rho = -0.65$ for the $L_{\mathrm{iso}}-\tau$ and
$\rho = -0.88$ for the $L_{X} - T_{\mathrm{brk}}$ relation) and have
surprisingly similar best-fit power law indices ($-1.19 \pm 0.17$
for $L_{\mathrm{iso}}-\tau$ and $-1.10 \pm 0.03$ for $L_{X} -
T_{\mathrm{brk}}$). Even more surprisingly, {we noted that
although $\tau$ and $T_{\mathrm{brk}}$ {represent} different GRB time
variables}, it appears that the first relation
($L_{\mathrm{iso}}-\tau$) extrapolates into the second one
for {timescales} $\tau \simeq T_{\mathrm{brk}}$. This fact
suggests that these two relations have a common origin, which we
conjecture to be kinematic. This relation adds to the
recently discovered relations between properties of the prompt and
afterglow GRB phases, indicating a much more intimate relation
between these two phases than hitherto considered.

\end{abstract}


\section{Introduction}

Gamma-ray Bursts (GRBs) are extremely bright explosions, with
isotropic luminosities exceeding $\sim 10^{54}$ erg/sec, durations
in the range $\sim 0.1 - 1000$ sec and energy of peak luminosity in
the $\gamma-$ray regime, $E_p \sim 1$ MeV, hence their name. They
are believed to originate in the collapse of stellar cores or the
mergers of neutron stars, processes which result in jet-like
relativistic outflows of Lorentz factors $\Gamma \sim 300$, whose
kinetic energy is converted efficiently into radiation at distances
$r \sim 10^{15} - 10^{18}$ cm at a relativistic blast wave (RBW) to
produce the observed events (for a review see Piran 2005). Following
their most luminous, prompt, $\gamma-$ray emission, they shift into
the afterglow phase with peak luminosity in the X-ray band and
duration of $\sim 10^5$ s, in which their localization can be
refined and optical detection can provide their redshift.

The theory of  RBW  slow-down indicated a smooth power law decay
$\propto t^{-1}$ for the flux of their afterglow X-ray light curves,
and indeed the pre-{\em Swift} sparsely sampled ones appeared
consistent with such a behavior. However, their more densely sampled
X-ray light curves with the XRT aboard {\em Swift} uncovered
significant deviations from this behavior. So, following the prompt
emission in $\gamma-$rays, the typical XRT afterglow consists
\citep{nousek} of a much steeper flux decline ($\propto t^{-3}\
\mbox{to}\ t^{-6}$), followed by a  $10^2 - 10^5$ sec period of
nearly constant flux, followed finally at $t = T_{brk}$ by the more
conventional power-law decline $\simeq t^{-1}$. In addition, {\em
Swift} follow-ups discovered also occasional flares on top of these
light curves, as late as $\sim 10^5$ sec since the Burst Alert
Telescope (BAT) trigger.

In their prompt phase GRBs exhibit a broad light curve diversity and
a large variance in their (estimated) RBW Lorentz factors ($\Gamma
\sim 100-1000$). These properties, along with the non-thermal
character of their spectra, suggest that at least in this phase,
GRBs are not likely to provide well defined, underlying systematics
that would allow a probe of the physics of prompt emission. However,
a number of correlations has been found between observables of the
prompt phase \citep{rie01,sch03,ghi06,ama09}, whose origin remains
still largely unaccounted for.


One of the first such correlations, made possible only after the
determination of the GRB redshifts by their afterglow emission, is
that between the burst peak isotropic luminosity, $L_{\rm iso}$, and
the spectral lag, $\tau$, between different energy bands in the GRB
spectrum. This relation has been studied in detail by a number of
authors
\citep{nor00,nor02,geh06,sch07,sta08,hak08,ukw10,ukw11,wan11} for
data sets obtained by different instruments aboard different
missions.  The general conclusion of all these studies has been an
anti-correlation $L_{\mathrm{iso}} \propto \tau^{-a}$ with a value
of $a \sim 0.77 - 1.8$. The spectral lag is defined as the
difference in the arrival time of high and low energy photons, and
is taken to be positive when the time of arrival of high energy
photons precedes that of low energy photons.

Normally the spectral lag is extracted between two arbitrary energy
bands in the observer frame and then corrected for the time
dilatation effect ($z-$correction) by multiplying the lag value in
the observer frame by $(1+z)^{-1}$. Moreover the observed energy
bands correspond to different energy bands at the GRB source frame
for different redshifts, and so one needs to take into account this
energy dependent factor ($k-$correction). Using the assumption that
the spectral lag is proportional to the pulse width which in turn is
proportional to the energy, \citet{geh06} approximately corrected
for this effect by multiplying the lag value in the observer frame
by $(1+z)^{0.33}$. Alternatively the $k-$correction can be done by
extracting the spectral lags in the GRB source frame. This is
accomplished by choosing two energy bands in the source frame and
then projecting these in the observer frame using the relation
$E_{\mathrm{obs}} = E_{\mathrm{source}}/(1 + z)$, such that the
projected energy bands lie in the \textit{Swift}  BAT energy range
$15-350$ keV. This alternative method of extracting the spectral
lags in the source frame rather than the observer frame, was
recently used by \citet{ukw11} to obtain a similar result for the
$L_{\mathrm{iso}}- \tau$ relation, but with a higher degree of
correlation between the variables.

The physical origin of the lag-luminosity relation is still unclear.
Some, including \citet{sal00} and  \citet{iok01}, attributed this to
a kinematic effect; \citet{sch04}, following an in depth analysis of
existing proposals, concludes in favor of the time evolution of the
emitting electrons, on the basis of a correlation between the energy
of the GRB peak emission and the burst instantaneous flux. However,
irrespective of its physical origin, this relation is a useful tool
in GRB science, not only for its use in distinguishing between long
and short bursts [with long bursts {in general} exhibiting larger
lags than short ones; {although it has been shown
\citep{geh06,hak07} that this may not be necessarily true for every
long GRB}], but also for its implementation, together with other
relations between GRB variables, in extending the Hubble diagram to
higher redshifts \citep{sch07,wan11}.

An altogether different correlation, pertaining to the GRB afterglow
phase, has been reported recently by \citet{dai10}. This work
presents a correlation between the X-ray luminosity, $L_X$, of the
plateau (or shallow decay) phase and the source frame break time
$T_{\rm brk}$ in the XRT light curves of long GRBs. Using a sample
of 62 \textit{Swift} long GRBs a correlation of the form $\log L_{X}
= \log a + b \log T_{\mathrm{brk}}$, with $\log a = 51.06\pm 1.02$
and $b = -1.06^{+0.27}_{-0.28}$ was obtained. A similar but steeper
correlation $(b = -1.72^{+0.22}_{-0.21})$ was also obtained for a
small group of GRBs which belong to the intermediate class (IC)
\citep{nor06} between short and long ones, indicating that these may
behave in a different way than the long GRBs. There have been claims
\citep{can11} that the Dainotti relation is just a selection effect
due to the flux detection limit for \textit{Swift}'s XRT which
prevents clear observation of faint light curves from high redshift
GRBs. This possibility was later investigated by \citet{dai11a} who
showed that there is no systematic bias against faint plateaus at
high $z$, thus confirming the existence of this relation. Moreover
\citet{dai11b} have obtained a number of significant correlations
between the afterglow phase X-ray luminosity parameter $L_{X}$ and
prompt emission parameters such as the isotropic energy
$E_{\mathrm{iso}}$, peak energy $E_{\mathrm{peak}}$ and the
variability parameter $V$ \citep{nor00}.\\

In this work we use a sample of 14 GRBs which are common in the
\citet{ukw10} and \citet{dai10} studies, to obtain and compare their
lag-luminosity and break time - X-ray luminosity relations in the
GRB source frame after doing the necessary $k$- and $z$-
corrections. The structure of the paper is as follows. In Section 2
we discuss briefly the correlations and computational procedures
involved in the spectral lag - isotropic luminosity relation of
\citet{ukw10} together with the X-ray luminosity of the GRB shallow
afterglow phase and its break time of \citet{dai10}. In Section 3 we
present our results and then in Section 4 we summarize our findings
and conclusions.

\section{GRB data}

In their work on the lag-luminosity relation \citet{ukw10} developed
a method for calculating the time-averaged spectral lag using a
modification of the cross-correlation function (CCF) with delay $d$
\citep{ban97} given by
\begin{equation}
\mbox{CCF}(d,x,y) = \frac{\sum_{i=\mathrm{max}(1,\,
1-d)}^{\mathrm{min}(N,\,
N-d)}\,x_{i}\,y_{i+d}}{\sqrt{\sum_{i}\,x_{i}^2\ \sum_{i}\,y_{i}^2}},
\label{ccf}
\end{equation}
where $x_{i}$ and $y_{i}$ are sets of time-sequenced data over $N$
bins, and then defining the spectral lag as the time delay which
corresponds to the global maximum of the CCF. {They obtained
the} uncertainty in the spectral lag using the Monte carlo method by
simulating 1000 light curves for each real light curve-pair and
calculating the CCF values using Equation (\ref{ccf}) for a series
of time delays. {Then they {obtained} the uncertainty from} the
standard deviation of the
CCF values per time delay bin of these simulated light curves.\\

To obtain the peak isotropic luminosity, $L_{\mathrm{iso}}$,
\citet{ukw10} fitted GRB spectra with the Band function
\citep{ban93} for the photon flux per unit photon energy using
\begin{equation}
N(E) = \left\{\begin{array}
    {l l}
A\left(\frac{E}{100\mathrm{keV}}\right)^{\alpha}\,e^{-(2+\alpha)E/E_{pk}},
& E\leq\left(\frac{\alpha - \beta}{2 + \alpha}\right)\,E_{pk} \\
A\left(\frac{E}{100\mathrm{keV}}\right)^{\beta}\left[\frac{(\alpha-\beta)E_{pk}}{(2+\alpha)100\mathrm{keV}}\right]^{\alpha-\beta}\,e^{(\beta
- \alpha)}, & \mbox{otherwise},
\end{array} \right. \label{band}
\end{equation}
where $A$ is the amplitude, $\alpha$ and $\beta$ are the low-energy
and high-energy spectral indices respectively, and $E_{pk}$ is the
peak energy of the $\nu F_{\nu}$ spectrum. The observed peak flux is
{expressed} in terms of the source frame spectrum $N(E)$
between energies $E_{1} = 1.0$ keV and $E_{2} = 10000$ keV  by
\begin{equation}
f_{\mathrm{obs}} = \int^{E_{2}/(1+z)}_{E_{1}/(1+z)}\,N(E)E\,dE.
\end{equation}
This was then used by \citet{ukw10} to compute the isotropic peak
luminosity from
\begin{equation}
L_{\mathrm{iso}} = 4\pi d_{L}^2 f_{\mathrm{obs}},
\end{equation}
where $d_{L}$ is the GRB luminosity distance computed in terms of
the redshift $z$ by
\begin{equation}
d_{L} = \frac{(1+z)c}{H_{0}}\int_{0}^{z}\frac{dz'}{\sqrt{\Omega_{M}
(1 + z')^3 + \Omega_{\Lambda}}}\label{luminosity},
\end{equation}
assuming a flat $\Lambda\mathrm{CDM}$ cosmological model with
$\Omega_{M} = 0.27$, $\Omega_{\Lambda} = 0.73$ and a Hubble constant
$H_{0}$ of $70 \; \mathrm{km \; s^{-1} \; Mpc^{-1}}$. The
uncertainty in $L_{\mathrm{iso}}$ {was} again determined
using Monte Carlo methods by calculating the luminosity for 1000
variations in the spectral parameters in (\ref{band}) for each GRB,
so that the real values and uncertainties are {given by} the sample means and
sample standard deviations respectively.

For the $T_{\mathrm{brk}} - L_{X}$ relation \citet{dai10} (see also
\citet{dai08}) used the fitting procedure of \citet{wil07} to
analyze the afterglow XRT light curves of a sample of \textit{Swift}
GRBs and derive the source frame parameters $T_{\mathrm{brk}}$ and
$L_{X}$ for each afterglow. The X-ray luminosity $L_{X}$ at the time
$T_{\mathrm{brk}}$, at which the light curve switches from the
plateau to the {declining} phase was calculated by using
\begin{equation}
L_{X} = \frac{4\pi d_{L}^2 F_{X}}{(1+z)^{1-\beta_{a}}},
\end{equation}
where $d_{L}$ is the same luminosity distance given by Equation
(\ref{luminosity}), $F_{X}$ is the observed flux at time
$T_{\mathrm{brk}}$, and $\beta_{a}$ is the spectral index obtained
for each afterglow \citep{eva09}. {Then they} computed the
uncertainties in the two parameters {by} using {a} Bayesian motivated
technique by \citet{dag05}.

\section{Results}

We collected a sample of 14 long GRBs (i.e. $T_{90} > 2s$) detected
by \textit{Swift} BAT between 2005-2008 with known redshifts ranging
from 0.703 (GRB 060904B)to 4.056 (GRB 060206), which {are}
common in the samples of \citet{ukw10} and \citet{dai10}
{studies}. The prompt and afterglow parameters of each GRB,
including the spectral lag, peak isotropic luminosity
$L_{\mathrm{iso}}$, peak energy $E_{\mathrm{pk}}$, break time
$T_{\mathrm{brk}}$, and X-ray luminosity $L_{X}$ are shown in Table
\ref{tbl-1}. The spectral lags are calculated between \textit{Swift}
energy bands $50-100 $ keV and $100-200$ keV in the GRB source
frame, after application of the $z$- and $k$- corrections, which are
obtained by multiplying the observed values by $(1+z)^{-0.67}$ as
described in the section above. Two of the GRBs (GRB 060206 and GRB
080603B) have negative spectral lags, meaning that the time of
arrival of low energy photons precedes that of high energy photons.
Although negative lags are not necessarily unphysical \citep{ryd05},
{we chose to exclude them due to the logarithmic nature of the lag
luminosity relation in our plots. This was also done in previous
studies of this relation by \citet{ukw10} and \citet{ukw11} and
others.}

\begin{figure}
\includegraphics[scale=1.5]{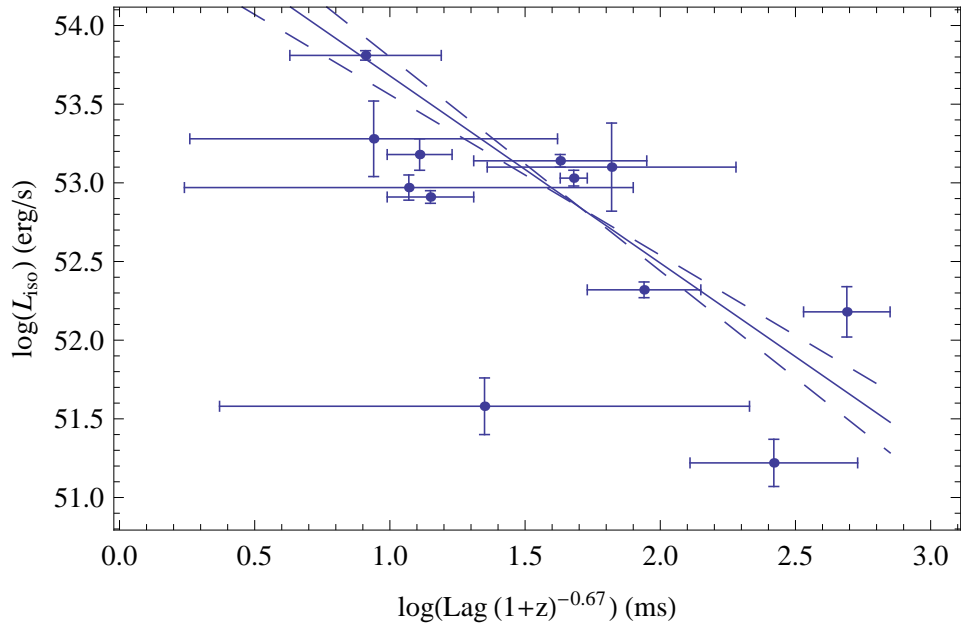}
\caption{Log-Log plot for the peak isotropic luminosity
$L_\mathrm{iso}$ vs source frame spectral lag between BAT channels
$(100-200 \mathrm{keV})$ and $(50-100 \mathrm{keV})$.\label{fig1}}
\end{figure}

We find that the $z$- and $k$- corrected spectral lag $\tau$ and the
peak isotropic luminosity $L_{\mathrm{iso}}$ are anti-correlated
with a correlation coefficient $\rho$ of $-0.65$, which is slightly
weaker than the value of $-0.73$ obtained by \citet{ukw10} for the
whole sample of 31 GRBs. Figure \ref{fig1} is a log-log plot of the
isotropic peak luminosity versus the $z$- and $k$-corrected spectral
lag with the following best-fit power-law curve\footnote{The
best-fit relations in this work were obtained by using the LinearFit
function available in the \textit{Experimental Data Analyst} package
in Mathematica.}
\begin{equation}
\log L_{\mathrm{iso}}(\mathrm{ergs/s}) = (54.87\pm0.29) - (1.19\pm0.17)\log(
(1+z)^{-0.67}\tau(\mathrm{ms})).\label{LT}
\end{equation}
The best-fit power-law index of $-1.19\pm0.17$ is consistent with
the earlier result $(-1.4\pm0.1)$ obtained by \citet{ukw10}  for the
full sample of 31 GRBs, with only redshift correction for the
spectral lags. Our result also agrees with the -1.14 power-law index
obtained by \citet{nor00} using spectral lags between the BATSE
energy bands $25-50$ keV and $100-300$ keV, and with those reported
by \citet{sta08} $(-1.16\pm0.21)$ and \citet{sch07}
$(-1.01\pm0.10)$.

\begin{figure}
\includegraphics[scale=1.5]{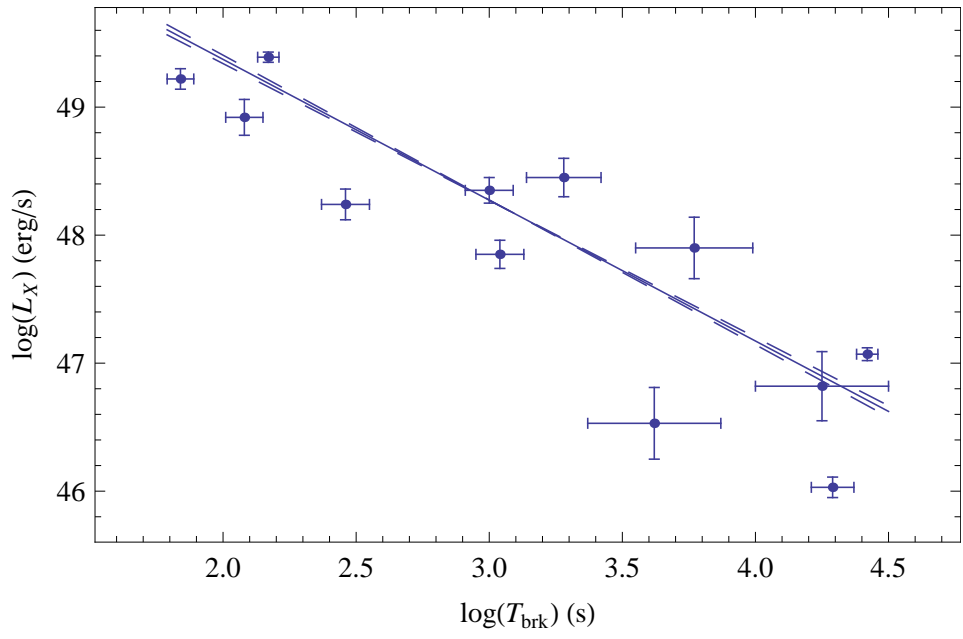}
\caption{Log-Log plot for the X-ray luminosity $L_{\mathrm{X}}$ vs
break time $T_{\mathrm{brk}}$.\label{fig2}}
\end{figure}

We have also obtained an anti-correlation with $\rho=-0.88$ between
the break time $T_{\mathrm{brk}}$ at the shallow-to-normal decay
transition in the GRB afterglow light curve and the X-ray luminosity
$L_{X}$. This is surprisingly stronger than the $\rho=-0.76$
 anti-correlation obtained by \citet{dai10} for the full sample of
 62 long GRBs. Figure \ref{fig2} shows a log-log plot of the break time
 $T_{\mathrm{brk}}$ versus the X-ray luminosity $L_{X}$ in the GRB
 source frame, with a fitted power-law relation given by
\begin{equation}
\log L_{X} (\mathrm{ergs/s}) = (51.57\pm0.10) -
(1.10\pm0.03)\log T_{\mathrm{brk}} (\mathrm{s}).\label{LxTa}
\end{equation}
{This} best fit power-law index is consistent with {the value}
$-1.06^{+0.27}_{-0.28}$ obtained by \citet{dai10} for the full
sample of GRBs. It also agrees with the value obtained by
\citet{str10} ($\sim -1.07$) for a small sample of 12 long GRBs, and
the recent study by \citet{qi10} who also obtained a power-law index
of $(-0.89\pm0.19)$ for a sample of 47 GRBs.

Noting the similarity of the slopes of {the} two relations in
(\ref{LT}) and (\ref{LxTa}), and the
fact that the ordinate of both is a luminosity while the abscissa
{represents} a time scale, we present in Figure \ref{fig3} a combined
plot of the two relations in a single figure. It is evident, quite
unexpectedly on our part, that one extrapolates into the other with
the combined relation given by
\begin{equation}
\log L (\mathrm{ergs/s}) = (54.69 \pm 0.06) - (1.07 \pm 0.014) \log T (\mathrm{ms}) .
\label{LtotT}
\end{equation}
The increased dynamic range provides for a much tighter relation {with
correlation coefficient} $\rho = -0.98$, and
with slope much closer to -1 than the individual relations.

\section{Discussion and Conclusion}

In the sections above we reviewed correlations between the
luminosities and time scales of two different stages in the GRB
development, namely the prompt emission and the shallow decay stage
of their afterglow. We then reproduced the correlations between $L -
\tau$ for the prompt emission and $L_X - T_{\rm brk}$ for the
afterglow, for the GRBs common in the data sets used by
\citet{ukw10} and \citet{dai10}, from the data {already}
present in the literature. {We have shown that although our
GRB sample is small, both relations are consistent (in terms of
power-law index and correlation coefficient) with the previous
relations obtained using larger samples. Moreover we have shown that
these two relations extrapolate very well into each other and give a
much tighter relation (Eq. (9)) than the individual relations
obtained so far.}

For the first time we also noted that although the relations in
Equations (\ref{LT}) and (\ref{LxTa}) represent different stages in
the GRB evolution, their power-law indices are surprisingly similar.
Yet from our data the source frame corrected spectral time lag and
break time $T_{\mathrm{brk}}$ do not appear to be correlated. The fact
that their normalizations are such that
they extrapolate into each other, suggests that prompt and afterglow
properties are interrelated. This fact could have been
actually surmised from the original lag treatment of \citet{nor00}
and the results of \citet{dai10}. The reader can easily confirm that
the relation of \citet{nor00} extrapolates into that of
\citet{dai10}. A correlation between
the prompt and afterglow phases has also been explored by
\citet{sal02} who obtained a correlation between the spectral lag
$\tau$ and rest frame jet-break time $\tau_{j}$ given by
\begin{equation}
\tau_{j} = 28^{+18}_{-11}\left(\frac{\tau}{1\mathrm{s}}\right)^{0.89
\pm 0.12}\ \mathrm{days}, \label{salmonson}
\end{equation}
using a sample of seven BATSE gamma-ray bursts. Another correlation
between prompt and afterglow quantities was recently obtained by
\citet{mar12}, who obtained a three-parameter correlation between
the rest frame isotropic energy in the prompt phase $E_{\gamma,\mathrm{iso}}$,
the peak of the prompt emission energy spectrum $E_{\mathrm{pk}}$,
and the X-ray energy emitted in the $0.3 - 10$ keV observed energy band
$E_{X,\mathrm{iso}}$ given by
\begin{equation}
E_{X,\mathrm{iso}} \sim \frac{E_{\gamma, \mathrm{iso}}}{E_{\mathrm{pk}}^{3/4}}.
\end{equation}
It was shown that this relation is robust and independent of the definition
of $E_{X,\mathrm{iso}}$. Moreover \citet{mar12} and \citet{ber12} showed that this
three-parameter relation is shared by both long and short gamma ray bursts
and {also} claim that the physical origin of such a relation is related to the
outflow Lorentz factor.

\begin{figure}
\includegraphics[scale=1.5]{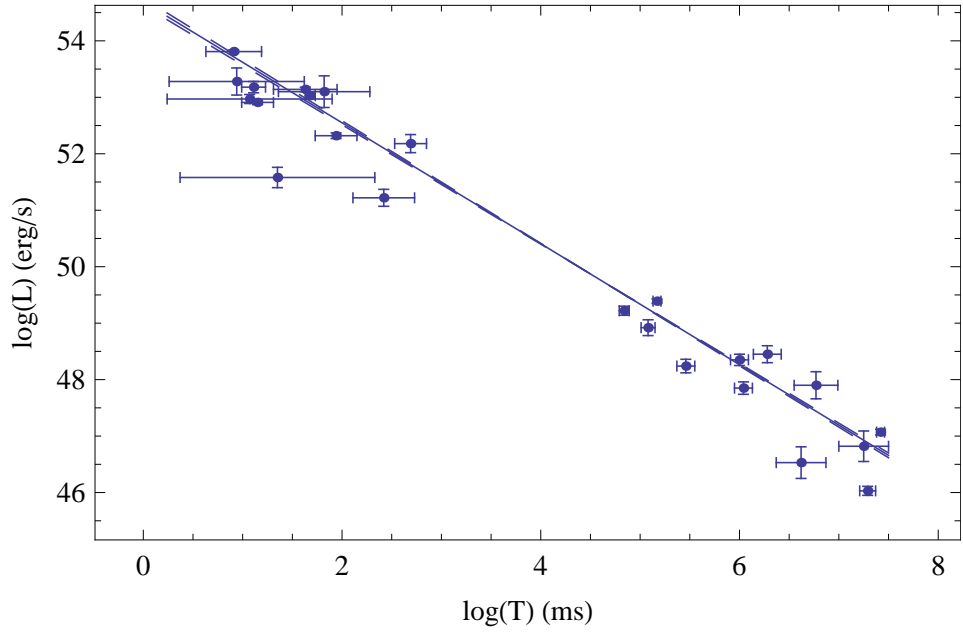}
\caption{Log-Log plot of the combined relations plotted in Figures
\ref{fig1} and \ref{fig2} above with $L_{\rm iso}$ or $L_X$ in the
ordinate and the lag $\tau$ or $T_{\rm brk}$, as appropriate, in the
abscissa. Apparently one relation extrapolates into the other.
\label{fig3}}
\end{figure}

At this stage we do not intend to speculate on the possible physics
underlying the correlations of (\ref{LT}) and (\ref{LxTa}). Instead, we present a brief review of
proposed explanations found in the literature.
Then we conclude that, if indeed the underlying physics is common, as
Figure \ref{fig3} suggests, the apparently common origin
of the two effects is basically kinematic.

In the case of the spectral lag-luminosity relation one possible
explanation for the observed lags involves the spectral evolution
during the prompt phase \citep{der98,koc03,ryd05} in which due to
cooling effects $E_{\mathrm{pk}}$ shifts towards a lower energy band
so that the temporal peak of the corresponding light curve will also
shift to lower energies, thereby resulting in the observed lag.
Another explanation for the spectral lag-luminosity relation is
based purely on kinematic effects
\citep{sal00,sal02,iok01,der04,she05,lu06}, where the peak
luminosity $L_{\mathrm{pk}}$ and spectral lag $\tau$ depend on a
single kinematic variable
\begin{equation}
D = \frac{1}{\Gamma(1 - \beta\cos\theta)(1 + z)}, \label{doppler}
\end{equation}
which represents the Doppler factor for ejecta in a jet, moving at
an angle $\theta$ from the line of sight with velocity $\beta \equiv
v/c$ at redshift $z$. This kinematic variable relates a proper
timescale $\tau$ in the GRB rest frame to the observed timescale $t$
given by
\begin{equation}
t = \frac{\tau}{D}, \label{timescales}
\end{equation}
so that if the spectral lag is due to some decay time scale
$\Delta\tau$ in the GRB restframe, then this will become $\Delta t =
\Delta\tau/D$ in the lab frame. Moreover assuming a power-law
spectrum with a low end of the form $\phi(E) \propto E^{-\alpha}$,
where $\alpha$ is the low energy spectral index, \citet{sal00}
showed that the peak luminosity varies as
\begin{equation}
L_{\mathrm{pk}} \propto D^{\alpha}, \label{lpk}
\end{equation}
with $\alpha \approx 1$ \citep{pre98}, so that equations
(\ref{timescales}) and (\ref{lpk}) lead to the lag luminosity
relation. The same argument was used by \citet{sal02} to explain
their correlation in equation (\ref{salmonson}), where in this case
the jet-break time $\tau_{j} \propto 1/D$.

In this approach, the dependence of luminosities and observed
timescales on the single variable $D$ leads to the conclusion that
the observed variety among GRBs has a kinematic origin, brought
through variation of the viewing angle $\theta$ or the Lorentz factor $\Gamma$
profile of the jet $\Gamma$, or both. So for example, \citet{sal00}
showed that the lag-luminosity relation is due only to a variation
in the line-of-sight $\Gamma$ among bursts, with high $\Gamma$
bursts having smaller spectral lags and low $\Gamma$ bursts exhibiting
longer ones.  On the other hand \citet{iok01} showed that
the lag-luminosity relation can be explained by variation in the
observer angle, $\theta_{v}$, from the axis of the jet, using a
simple jet in which $\Gamma = \mathrm{const.}$ for $\theta <
\theta_{j}$, and zero emission for $\theta > \theta_{j}$, where
$\theta_{j}$ is the opening angle of the jet. In this case the lags
arise due to the path difference between the near and far edges of
the emitting region such that bright (dim) bursts with short (long)
spectral lags correspond to small (large) viewing angle.

An explanation for the anticorrelation between the duration of the
intrinsic plateau phase of the GRB light curve and X-Ray luminosity
has been proposed by \citet{dal11} using a model in which energy
from a long-lived central engine is continuously injected to balance
the radiative losses. These radiative losses will be stronger for
higher luminosity, thus leading to shorter plateaus. Another
explanation which is  based on the kinematic effect discussed above
was proposed by \citet{eic06}, who claimed that the flat (or sometimes
slightly rising) decay phase
of the afterglow lightcurve results from the combination of the
decaying tail of the prompt emission and early afterglow observed
at viewing angles slightly outside the edge of the jet. For such
``offset'' viewing angles the afterglow flux initially rises at
early times when the beaming of radiation away from the line of
sight gradually decreases, then rounds off as the beaming cone
expands to include the line of sight, and finally joins the familiar
decaying light curve.

Clearly, the relations given by equations (\ref{LT}), (\ref{LxTa})
and (\ref{LtotT}) call for further analysis with larger data sets to
determine whether the indices and normalization of these relations
are indeed consistent with those presented above. Since the
relations {in} (\ref{LT}), (\ref{LxTa}) correspond to {the} prompt
and afterglow phases of the GRB evolution, the similarity of their
power-law indices and normalizations (they extrapolate into each other in Figure (\ref{fig3}))
is an indication that a common process, probably kinematic, is
responsible for the observed spectral lags and the shallow decay
phase of the afterglow light curve. As discussed above,
both these relations were attributed individually \citep{iok01,eic06} to the same kinematic
process, namely viewing the grb jets at ``off-beam'' lines of sight.
The results presented in this
paper are in accordance with this explanation.

\acknowledgments
We would like to acknowledge useful discussions with Takanori Sakamoto.
J.S. gratefully acknowledges financial support from the University
of Malta during his visit at NASA-GSFC.

\clearpage

\clearpage

\clearpage

\clearpage

\begin{deluxetable}{lrrrrrr}
\tabletypesize{\scriptsize} \tablecaption{GRB redshift, prompt and
afterglow parameters taken from \citet{ukw10} and
\citet{dai10}.\label{tbl-1}} \tablewidth{0pt} \tablehead{
\colhead{GRB} & \colhead{Redshift} & \colhead{$E_{\mathrm{pk}}\
(\mathrm{keV})$\tablenotemark{a}} & \colhead{$L_{\mathrm{iso}}\
(\mathrm{erg/s})$} & \colhead{Lag (ms)} & \colhead{$\log L_{X}\
(\mathrm{erg/s})$} & \colhead{$\log T_{\mathrm{brk}}\ (\mathrm{s})$}
} \startdata
GRB050401 & 2.899 & $119^{+16}_{-16}$ & $(1.38^{+0.16}_{-0.13})\times10^{53}$  & $106\pm118$ & $48.45\pm0.15$ & $3.28\pm0.14$ \\
GRB050603 & 2.821 & $349^{+18}_{-18}$ & $(6.32^{+0.47}_{-0.34})\times10^{53}$ & $20\pm18$ & $46.82\pm0.27$  & $4.25\pm0.25$ \\
GRB050922C & 2.199 & $[133^{+468}_{-39}]$  & $(5.17^{+28.00}_{-0.01})\times10^{52}$  & $19\pm72$ & $48.92\pm0.14$  & $2.08\pm0.07$\\
GRB060206 & 4.056 & $75^{+12}_{-12}$ & $(6.28^{+2.50}_{-0.62})\times10^{52}$ & $-163\pm189$ & $48.65\pm0.14$ & $3.15\pm0.10$ \\
GRB060210 & 3.913 & $207^{+66}_{-47}$ & $(8.53^{+2.75}_{-0.92})\times10^{52}$ & $34\pm195$ & $47.90\pm0.24$  & $3.77\pm0.22$ \\
GRB060418 & 1.490 & $230^{+23}_{-23}$  & $(1.96^{+0.43}_{-0.13})\times10^{52}$  & $162\pm101$ & $47.85\pm0.11$ & $3.04\pm0.09$ \\
GRB060904B & 0.703 & $103^{+59}_{-26}$ & $(2.18^{+3.59}_{-0.32})\times10^{51}$ & $32\pm273$ & $46.53\pm0.28$ & $3.62\pm0.25$ \\
GRB060908 & 1.884 & $124^{+48}_{-24}$ & $(1.54^{+22.50}_{-0.22})\times10^{52}$ & $134\pm253$ & $48.24\pm0.12$ & $2.46\pm0.09$ \\
GRB061007 & 1.262 & $498^{+34}_{-30}$ & $(1.01^{+0.20}_{-0.08})\times10^{53}$ & $82\pm9$ & $49.39\pm0.04$ & $2.17\pm0.04$ \\
GRB061121 & 1.315 & $606^{+56}_{-45}$ & $(7.89^{+1.02}_{-0.47})\times10^{52}$ & $25\pm11$ & $48.35\pm0.10$ & $3.00\pm0.09$ \\
GRB070306 & 1.496 & $[76^{+131}_{-52}]$ & $(8.67^{+13.50}_{-0.27})\times10^{51}$ & $900\pm408$ & $47.07\pm0.05$ & $4.42\pm0.04$ \\
GRB071020 & 2.145 & $322^{+50}_{-33}$ & $(1.27^{+0.64}_{-0.15})\times10^{53}$ & $28\pm9$ & $49.22\pm0.08$ & $1.84\pm0.05$ \\
GRB080430 & 0.767 & $[67^{+85}_{-51}]$ & $(1.03^{+1.30}_{-0.07})\times10^{51}$ & $388\pm397$ & $46.03\pm0.08$ & $4.29\pm0.08$ \\
GRB080603B & 2.689 & $71^{+10}_{-10}$ & $(2.99^{+1.25}_{-0.30})\times10^{52}$ & $-172\pm56$ & $48.88\pm0.29$ & $2.92\pm0.24$\\
\enddata
\tablenotetext{a}{Values in brackets represent estimated values
obtained using the method in \citet{sak09}}
\end{deluxetable}

\end{document}